 \documentclass[5p,twocolumn,10pt,authoryear]{elsarticle}
\pdfoutput=1
\usepackage[utf8x]{inputenc}
\usepackage[english]{babel}

\usepackage{url}

\usepackage{graphicx}
\usepackage{booktabs}

\usepackage{amsmath}
\usepackage{amsfonts}
\usepackage{amssymb}

\begin{document}

\begin{frontmatter}

\title{Revealing cell assemblies at multiple levels of granularity}

\author[label1]{Yazan N. Billeh\fnref{f1}\corref{cor1}}

\fntext[f1]{Email: ybilleh@caltech.edu}
\address[label1]{Computation and Neural Systems Program, California Institute of Technology, Pasadena, California 91125}
\cortext[cor1]{denotes equal contribution}

\author[label2]{Michael T. Schaub\fnref{f2}\corref{cor1}}
\fntext[f2]{Email: michael.schaub09@imperial.ac.uk}
\address[label2]{Department of Mathematics, Imperial College London, London SW7 2AZ, U.K.}

\author[label3]{Costas A. Anastassiou}
\address[label3]{Allen Institute for Brain Science, Seattle, Washington 98103}

\author[label2]{Mauricio Barahona}

\author[label3]{Christof Koch}

\begin{abstract}

{\it Background:} Current neuronal monitoring techniques, such as calcium imaging and multi-electrode arrays, enable recordings of spiking activity from hundreds of neurons simultaneously. Of primary importance in systems neuroscience is the identification of cell assemblies: groups of neurons that cooperate in some form within the recorded population. 

{\it New Method:}
We introduce a simple, integrated framework for the detection of cell-assemblies from spiking data without {\it a priori} assumptions about the size or number of groups present. We define a biophysically-inspired measure to extract a directed functional connectivity matrix between both excitatory and inhibitory neurons based on their spiking history. The resulting network representation is analyzed using the Markov Stability framework, a graph theoretical method for community detection across scales, to reveal groups of neurons that are significantly related in the recorded time-series at different levels of granularity. 

{\it Results and comparison with existing methods:}
Using synthetic spike-trains, including simulated data from leaky-integrate-and-fire networks, our method is able to identify important patterns in the data such as hierarchical structure that are missed by other standard methods. We further apply the method to experimental data from retinal ganglion cells of mouse and salamander, in which we identify cell-groups that correspond to known functional types, and to hippocampal recordings from rats exploring a linear track, where we detect place cells with high fidelity.

{\it Conclusions:}
We present a versatile method to detect neural assemblies in spiking data applicable across a spectrum of relevant scales that contributes to understanding spatio-temporal information gathered from systems neuroscience experiments.
\end{abstract}

\end{frontmatter}

\section{Introduction}
As capabilities for parallel recordings from large neuronal populations continue to improve \citep{Ahrens2013,Buzsaki2004} experimentalists are now able to probe neural population encoding in ever more detail. 
These experimental advances allow the study of the intricate links between topology and dynamics of neural interactions, which underpin the functional relationships within neural populations.
One such example is the activity of cell assemblies.
The problem is to identify groups of neurons (termed cell assemblies) within a large number of simultaneously recorded neurons where, due to functional cooperativity, each cell in an assembly is more similar in its temporal firing behavior to members of its own group than to members of other groups.
Such strongly intertwined activity patterns are believed to underpin a wide range of cognitive functions \citep{Hebb1949,Harris2005,Buzsaki2010}.
However, the reliable identification of cell assemblies remains challenging.

Here we introduce a technique to identify such neuron assemblies directly from multivariate spiking data, based on two steps:   the definition of a simple biophysically-inspired similarity measure obtained from the observed spiking dynamics, followed by its analysis using a recent framework for multiscale community detection in weighted, directed graphs.
A variety of techniques have been proposed to cluster spike-train groups to date, and have shown promising results in particular settings \citep{Fellous2004,Feldt2009,Humphries2011,Lopes-dos-Santos2011,Lopes-Dos-Santos2013,Quiroga2009,Abeles2001,Laubach1999,Peyrache2010,Gansel2012}.
In contrast to these techniques, our methodology provides a dynamics-based framework, in which both the similarity measure and the community detection method are geared towards incorporating key features of neural network dynamics.
The framework is purposely designed to be simple, yet capturing a breadth of features not present concurrently in other methods.

Our similarity measure evaluates the association between neuron pairs based on their spiking history and integrates three features that are key for a network-based analysis of neuro-physiological data: 
(i) an intuitive biophysical picture, allowing a simple interpretation of the computed associations; 
(ii) a measure that is directed in time, hence asymmetric in the sense that spike-time dependent information is retained (\textit{e.g.},  spiking of neuron A precedes that of neuron B);
(iii)~excitatory and inhibitory interactions are both included yet treated differently, inspired by their distinct effects on post-synaptic cells.

The detected dynamic associations are interpreted as an induced functional network, which is used to identify neuronal assemblies using a directed version of the recently introduced Markov Stability framework for community detection in graphs \citep{Delvenne2010}. Unlike other approaches, this framework allows us to analyze directed networks and search for cell assemblies at all levels of granularity, from fine to coarse levels of resolution, extracting relevant, possibly hierarchical groupings in spike trains without \textit{a priori} assumptions about the groups present.
In the following, we present our framework and evaluate it on a series of examples, including synthetic spike-trains and leaky-integrate-and-fire network models. We also apply it to experimental datasets from retinal ganglion cells and hippocampal pyramidal neurons.

\section{Materials and Methods}

Most existing methods to detect groups in spike-train neuronal population data are based on the following generic paradigm \citep{Fellous2004,Feldt2009,Humphries2011,Lopes-dos-Santos2011}.
First, a metric is defined to quantify the relationship between all neuron pairs leading to a $N\times N$ association matrix, where $N$ is the number of observed neurons.
We call this the \textit{functional connectivity matrix} (FCM) hereafter.
Every $(i,j)$ entry in this matrix is a non-negative number that indicates how similar the spike trains of neurons $i$ and $j$ are over the observed time.
Second, the FCM is clustered, \textit{i.e.}, partitioned into different groups \citep{Newman2004,Fortunato2010,Aggarwal2014}.

Here we introduce a simple framework that addresses both of these steps in a consistent and integrated manner, focusing on the dynamical relations between neurons: a new directed (`causal')  biophysically-inspired measure is introduced to calculate the FCM, which is then analyzed using the recently introduced dynamics-based technique of Markov Stability for community detection \citep{Delvenne2010,Lambiotte2009,Schaub2012,Delvenne2013} to identify cell assemblies at multiple scales
in the neuronal population.

The numerics are performed in MATLAB (2011b or later versions). Code implementing the algorithm for spike-train analysis is available upon request and will be made available at github.com/CellAssembly/Detection.

\subsection{Biophysically-inspired causal measure of spike-train similarity}
\begin{figure}[tb!]
  \centering
  \includegraphics{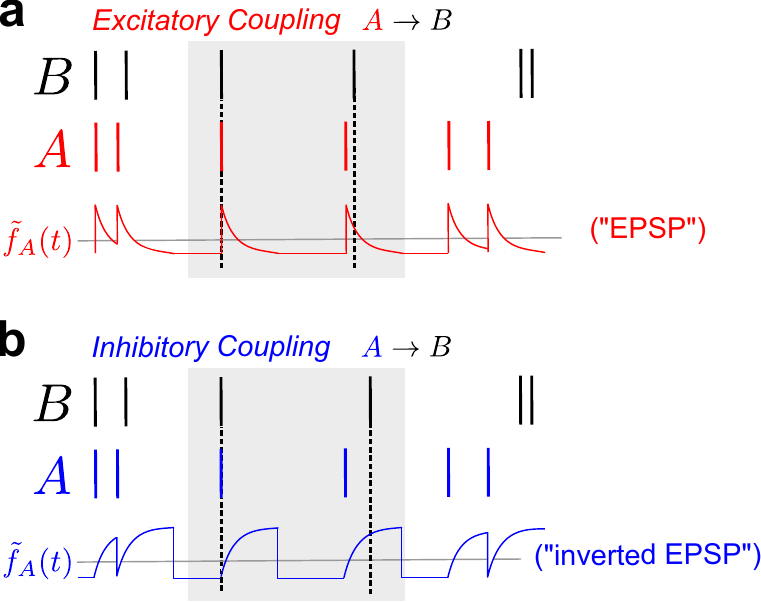}
  \caption{Biophysically-inspired measure of spike-train similarity leading to functional coupling between neurons. Quantification of the coupling induced by:   \textbf{(a)} excitatory neuron $A$ on neuron $B$ and \textbf{(b)} inhibitory neuron $A$ on neuron $B$. Note that both profiles shown are normalized so that the signal has zero mean (see text).}
  \label{fig:1}
\end{figure}

A plethora of metrics exists to describe the relationship between two signals, ranging from generic measures, such as cosine similarity and Pearson or Spearman correlation coefficients, to specialized measures designed for spike-train analysis \citep{Kreuz2013,Lyttle2011,Victor1996,Fellous2004,Schreiber2003,Rossum2001,Okatan2005,Vincent2012}.
Although these methods can be well suited in particular contexts, they only partially account for three important features for network-driven analyses of neural recordings.
First, most current metrics are based on statistical arguments lacking a simple biophysical interpretation that would allow the use of relevant biophysical characteristics of neuronal dynamics.
Second, most commonly used measures are distance metrics, \textit{i.e.,} symmetric by construction, and thus neglect spike-timing information contained in the ordering of events.
Finally, to the best of our knowledge, all measures ignore whether the neurons under consideration are excitatory or inhibitory.
While an even finer characterization of neuronal subtypes could be of further interest, the distinction between excitatory and inhibitory neurons underpins fundamental balances in neuronal network dynamics and should be reflected in the analysis of data.
Here, we propose a similarity measure that incorporates these three ingredients in a simple, intuitive form (see Figure \ref{fig:1}).

Consider first an excitatory neuron A connected to neuron B.
The action potentials of A induce excitatory postsynaptic potentials (EPSPs) in neuron B,  increasing the likelihood of neuron B firing.
These EPSPs can be, to a first approximation, modeled by an exponentially decaying time profile

$$\xi_\text{exc}(t) = e^{-t/\tau}$$
with synaptic time constant $\tau$. 
Since detailed information about synaptic weights and membrane potentials is unavailable in neuronal population experiments, we adopt a simple strategy to compute the coupling strength $S_{AB}$ from the observed spiking data.
The general idea is that for each spiking event of neuron B  (at time $t_{i}^B$), we propagate a `virtual' EPSP  from the immediately preceding spike of neuron A (at time $t_{i}^A$).
We then compute all such contributions that neuron A  would have made to the membrane potential of neuron B at each of its spikes, and sum them appropriately discounting spurious effects.

More precisely, we obtain the functional connectivity from neuron A to neuron B as follows:\\
\textbf{(i)} Define the signal $f_{A}(t)$ that reflects the (virtual) influence of neuron A onto a potential firing event at any other neuron taking place at time $t$:

\begin{equation}
  f_{A}(t) = \xi_\text{exc}(t- t_\text{last}^A) =  e^{-(t-t_\text{last}^A)/\tau},
\end{equation}

where $t_\text{last}^A = \max_i(t^A_i | t^A_i \leq t), i=1,\ldots, N_A$, is the time of the last preceding spike of neuron A (if there is no such spike we set ${t_\text{last}^A=-\infty}$). \\

\textbf{(ii)} It then follows that all contributions from neuron A to B can be written as the sum
$\sum_{i=1}^{N_B} f_{A}(t^B_i).$
To gain some intuition, note that every time B fires a spike, the potential contribution to this spike by neuron A is computed by summing the values that $f_A(t)$ takes at the times of B firing, $t^B_i$. 
If neuron B always fires shortly after A spikes, the sum  $\sum_i^{N_B} f_A(t_i^B)$ will be large. 
If neuron B fires after A but with some delay (e.g., because an integration with other neurons is required), this sum will be smaller.  If neuron B never fires shortly after A, this sum will be zero.

To discount spurious correlation effects, we center and normalize the signal $f_A(t)$ first to obtain the new signal $\tilde{f}_A(t)$, which has zero mean and peak amplitude one (Figure \ref{fig:1}a)
\begin{equation}
  \tilde{f}_A(t) = \dfrac{f_{A}(t) - \langle f_{A}\rangle}{1-\langle f_{A}\rangle},
\end{equation} 
where $\langle f_{A}\rangle = \frac{1}{T}\int_0^T f_{A}(t) dt \leq 1$ is the mean over the recorded time.
We then compute the effective coupling:
\begin{equation}
F_{AB} = \dfrac{1}{N_{AB}}\sum_{i=1}^{N_B}\tilde{f}_{A}(t^B_i),
\end{equation} 
and we have additionally divided by $N_{AB}= \max(N_A,N_B)$ to guarantee that the maximal coupling $F_{AB}$ (between two identically firing neurons) is normalized to $1$.
The coupling between neuron A and B is then defined as the thresholded value:

\begin{equation}
\label{eq:similarity}
 S_{AB} = \max({F_{AB}, 0}).
\end{equation}

From this definition, it follows that if an action potential from neuron A is always closely followed by a spike from neuron B, this will correspond to a strong coupling $S_{AB}$ between these neurons. Note that, in addition to being biophysically inspired, the defined measure~\eqref{eq:similarity} is non-symmetric ($S_{AB} \neq S_{BA}$).

Suppose that neuron A is known to be inhibitory. The coupling strength from neuron A onto another neuron is obtained following a similar approach (Figure~\ref{fig:1}b), yet recognizing that inhibitory post-synaptic potentials (IPSPs)  decrease the likelihood of firing.
To reflect this influence, we adopt an `inverted' exponential profile 
$$\xi_\text{inh}(t) = 1-e^{-t/\tau},$$
truncated when it reaches 99\% of its steady state value.
Hence, if neuron B always fires shortly after the firing of the inhibitory neuron A, it will accumulate a negative dependence from which we deduce that there is no significant inhibitory functional relation between these neurons. 

The time scale $\tau$ is a parameter inspired by synaptic time constants, and can thus be adapted to reflect prior information about the recorded neurons.
Although more sophisticated schemes to estimate or tune this parameter are certainly possible (\textit{e.g.}, choosing a different $\tau_\text{exc}$ for excitatory and $\tau_\text{ inh}$ for inhibitory neurons), here we follow the simplest choice 
$\tau_\text{exc} = \tau_\text{ inh} = \tau$ throughout. The method is robust to the choice of $\tau$: we have used $\tau=5$ ms for the experimental data and $\tau=3$ ms for the leaky-integrate-and-fire (LIF) simulation data, and have checked that the results remain broadly unaltered for values of $\tau$ in this range.

The main aim of our measure is simplicity, flexibility and generality, while retaining the key biophysical features outlined above.
Because of its generality, highly specialized measures of spike-train associations could be tuned to outperform our simple measure for particular examples.
However, it is often unknown beforehand what features of the data are of importance for the analysis.
Hence having such a flexible measure allows for a broad search for structure in recorded data.
Once a hypothesis is formed, or particular aspects need to be investigated in more detail, more specialized association metrics could be used in conjunction with the community detection algorithm presented below.
In the absence of knowledge about the specific cell types of experimentally recorded neurons we obtain the FCM using the excitatory metric.
Already today, however, there are means to separate cell types (e.g. fast spiking interneurons) based on their electrophysiological signature \citep{Bartho2004} and with the advancement of optical physiology and genetic tools, additional information about the cell types of the recorded cells is becoming more routine.
Hence it will be possible in the future to use specialized coupling functions (instead of exponential) depending on the neuronal sub-type recorded.

\subsection{Markov Stability for community detection at all scales}
The Markov Stability method is a versatile, dynamics-based tool for multiscale community detection in networks without \textit{a priori} assumptions about the number or size of the communities  \citep{Delvenne2010,Lambiotte2009,Schaub2012,Delvenne2013}.
Here we extend the use of Markov Stability to directed networks to find coherent groupings of neurons in the FCM created from the observed spiking data.
Under our framework, we interpret the FCM as a directed network, and the graph communities revealed by our analysis correspond to groups of neurons with strong excitatory and/or inhibitory couplings extracted from the dynamics. 
Therefore the graph partitioning problem solved using the Markov Stability method is linked to the detection of putative cell assemblies, \textit{i.e.}, groups of neurons with a strong dynamical influence on each other.

\begin{figure}[tb!]
  \centering
  \includegraphics{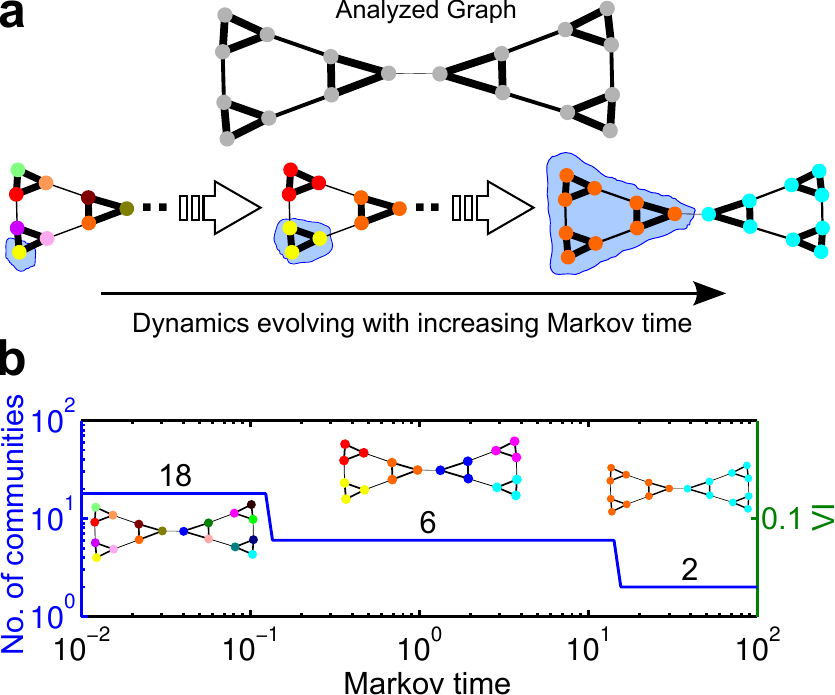}
  \caption{Schematic of Markov Stability method used to partition the functional network.
  \textbf{(a)} A diffusion process on a network can be used to reveal the structure of a graph. 
As the diffusion explores larger areas of the network, it enables the Markov Stability method to scan across all scales and reveal relevant partitions at different levels of granularity. 
  \textbf{(b)}  The graph analyzed has a pre-defined multi-scale community structure, given by a hierarchy of triangles.
The number of communities found are plotted as a function of the Markov time (see \textbf{(a)})long plateaus indicate well-defined partitions into 18 nodes (each node on its own), six communities (small triangular structures), and two communities (aggregated, larger triangles).
Note that in this example, the variation of information (VI) is zero for all Markov times, indicating that all three partitions are relevant at different levels of resolution.}
  \label{fig:2}
\end{figure}

The main notion underpinning the Markov Stability method is the intimate relationship between structure and dynamics on a graph. 
A dynamics confined to the topology of a network can uncover structural features of the graph by observing how a dynamical process, such as a simple diffusion, unfolds over time. 
In particular, if the graph contains well defined substructures, such subgraphs will trap the diffusion flow over a significantly longer time than expected if it were to happen on an unstructured graph.
This idea is readily illustrated by the example of ink diffusing in a container filled with water.
If the container has no structure, the ink diffuses isotropically.
If the container is compartmentalized, the ink would get transiently trapped in certain regions for longer times until it eventually becomes evenly distributed throughout.
In a similar manner, by observing the dynamics of a diffusion process we can gain valuable information about the structural organization of the graph (Figure \ref{fig:2}).
We use this concept to define a cost function to detect significant partitions in the graph, as follows. 

To make these notions precise, consider a network with a Laplacian matrix $L=D-A$, where $A$ is the weighted adjacency matrix ($A_{ij}$ is the weight of the \emph{directed} link from node $i$ to node $j$) and $D=\text{diag} (A\mathbf{1})$ is the diagonal out-degree matrix ($\mathbf{1}$ is the vector of ones).
For ease of explanation, consider first a strongly connected graph,  \textit{i.e.}, we can traverse the graph along its directed edges such that every node can be reached from any other node. 
On such a network, let us define a continuous diffusion process:
\begin{equation}
  \mathbf{\dot p} = -\mathbf{p} \, D^{-1}L,
  \label{eq:diffusion}
\end{equation}
where $\mathbf{p}$ is the $1 \times N$ probability vector describing the probability of a random walker to visit different nodes over time.
Note that the probability vector remains properly normalized: $\mathbf{1}^T\mathbf{p} = 1$ at all times. 
For an undirected connected graph, this dynamics converges to a unique stationary distribution 
$\boldsymbol{\pi} = \boldsymbol{d/(d^T1)}$.
For \textit{directed} graphs the stationary distribution has to be computed by solving $\mathbf{\dot p} =0$, \textit{i.e.}, it corresponds to the dominant left eigenvector of $D^{-1}L$.
If the graph is not strongly connected (\textit{e.g.},  if it contains a sink), the diffusion process~\eqref{eq:diffusion} is generalized to include the standard random `teleportation' term inspired by Google's page-rank algorithm~\citep{Brin1998,Lambiotte2009,Lambiotte2012}:  the random walker is transported from any node to a random node in the graph with a small, uniform probability $\alpha$ (set here to the commonly adopted value $\alpha = 0.15$), while in the case of a sink node, it will be teleported with unit probability.
This term guarantees that the process is ergodic with a unique stationary probability distribution.

Consider a partition of this network encoded in a $N \times c$ indicator matrix $H$, with $H_{ij}=1$ if node $i$ belongs to community $j$. We then define the Markov Stability of the partition $r(t_M,H)$, as the probability that a random walker at stationarity starts in community $i$ and ends up in the same community after time $t_M$ minus the probability of such an event happening by chance, summed over all communities and nodes.  In matrix terms, this may be expressed as:
\begin{equation*}
	S(t_M) = \Pi \exp(-t_M D^{-1}L) - \boldsymbol{\pi}\boldsymbol{\pi}^T
\end{equation*}
\begin{equation*}
	r(t_M,H) = \text{trace}\left [H^T S(t_M) H\right],
\end{equation*}
where $\Pi = \text{diag}(\boldsymbol{\pi})$ and $t_M$ denotes the Markov time describing the evolution of the diffusion process.
Finding a good partition (or clustering) requires the maximization of the Markov Stability in the space of possible graph partitions for a given $t_M$, an optimization that can be carried out with a variety of optimization heuristics. Here we use a locally greedy optimization, the so-called Louvain algorithm, which is highly efficient~\citep{Blondel2008}. 
In order to deal with the fact that $S(t_M)$ is in general asymmetric due to the directed nature of the graph, we use the directed notion of Markov Stability and use the Louvain algorithm to optimize 
$H^T \frac{1}{2}(S+S^T)H$, 
which is mathematically identical to optimizing $r(t_M,H)$, i.e., we still consider the \textit{directed} network.

Our algorithm then scans across all Markov times to find the set of relevant partitions at different Markov times. With increasing Markov time, the diffusion explores larger regions of the network, resulting in a sequence of increasingly coarser partitions, each existing over a particular Markov time scale. The Markov time may thus be interpreted as a resolution (or \emph{granularity}) parameter, and, as we sweep across resolutions, we detect communities at different levels of granularity without imposing a particular resolution \textit{a priori}.
This dynamic sweeping~\citep{Schaub2012} allows us to detect assemblies of different sizes and even hierarchical structures that would potentially go undetected if we were to use a method with a fixed intrinsic scale~\citep{Newman2004a,Fellous2004,Fortunato2007,Feldt2009,Humphries2011,Lopes-dos-Santos2011}.
It is important to remark that the Markov time $t_M$ used for the diffusive exploration of the network is not to be confused with the physical time of the spike-train dynamics.
We remark that the time constant $\tau$ of our similarity measure is not related to the Markov time in general. The Markov time is used here as a tool to uncover the different scales in the data and should thus be seen as distinct from the biophysical (real) time.

To select meaningful partitions across levels of granularity, we use two measures of robustness. 
Firstly, a relevant partition should be persistent over a long Markov time-horizon, \textit{i.e.}, it should be robust with respect to the change in Markov time and thus lead to an extended plateau in Markov time. Secondly, a relevant partition should be consistently found by the optimization algorithm, \textit{i.e.}, it should be robust to random initializations of the Louvain optimization.  
In order to establish the optimization robustness, we run the Louvain algorithm 100--500 times per Markov time and compare the partitions obtained by means of the variation of information (VI) distance metric~\citep{Meila2003,Meila2007}.
The variation of information can be thought of as an information-theoretic distance between two partitions that is naturally invariant to a relabeling of the groups and which has proved useful as a standard tool to compare partitions in the context of community detection~\citep{Fortunato2010}.
The normalized VI between two partitions $\mathcal P^\alpha$ and $\mathcal P^\beta$ is defined as~\citep{Meila2007}:
\begin{equation}
\label{eq:VI}
 \text{VI}(\mathcal P^\alpha, \mathcal P^\beta) = \frac{2 \,  H(\mathcal P^\alpha, \mathcal P^\beta) - H(\mathcal P^\alpha) - H(\mathcal P^\beta)}{\log N},
\end{equation}
where $H(\mathcal P) = -\sum_{\mathcal C} p(\mathcal C) \log p(\mathcal C)$ is the Shannon entropy of the relative frequency $p(\mathcal C) = n_{\mathcal C}/N$ of a node belonging to community $\mathcal C$ in a partition $\mathcal P$ and $H(\mathcal P^\alpha, \mathcal P^\beta)$ is the Shannon entropy of the corresponding joint probability.
We then calculate the average variation of information ($VI$) over all pairs in the ensemble of solutions from the optimization.
When $VI \approx 0$, the solutions obtained by the different optimizations are very similar to each other indicating a robust partitioning. 
When $VI \approx 1$ each run of the optimization obtains a different partition, indicating a non-robust clustering.  Such clear-cut communities are not always found. However, we have shown \citep{Schaub2012,Delmotte2011} that sudden drops and dips in the $VI$ are indicative of a clustering becoming more robust than expected for its average community size. In realistic datasets, we thus search for partitions with a long Markov time plateau and a low value (or a pronounced dip) of $VI$ as the criterion to find meaningful partitions.
An illustration of the Markov Stability framework is displayed in Figure \ref{fig:2}b, where we exemplify how the graph community structure can be detected at different scales without \textit{a priori} assumptions about the number of communities.  Furthermore, our scanning across all Markov times allows for the detection of the appropriate scale for community detection, without imposing \textit{a priori} a particular scale that might not be relevant to the analyzed data, as is implicitly done in other methods \citep{Schaub2012}.

\subsection{Synthetic spiking data}
To assess the capabilities of the framework, we generated synthetic spiking datasets with realistic statistical properties resembling those observed in experiments, yet with added temporal structure.

\subsubsection{Synthetic data with embedded and hierarchical cell assemblies}
Surrogate spike-train data were created from groups of units with variable sizes.
Each group $G_i$ was assigned a firing rate ($f_i$) and a level of jitter ($J_i$).
The firing times of each \emph{group} were drawn from a uniform distribution according to the specified firing frequency $f_i$, and the firing times for each \emph{unit} were chosen from a uniform distribution with a range $\pm J_i$ around the group firing time.
To account for refractory periods, we resampled if the resulting spike time conflicted with the refractory period of the unit.
We used a similar scheme to generate synthetic spiking data with a hierarchical structure, but in this case each group was divided into two subgroups: units within each subgroup always fire together, whereas between two subgroups the firing window was aligned only every second time.
As before, the firing times of the individual groups were chosen randomly from a uniform 
distribution and were not correlated in time. This firing pattern establishes a two-level hierarchical relation between the individual units.

\subsubsection{Synthetic data with feedforward-like firing patterns}
Synthetic spiking patterns that emulate the activity of feedforward networks were created from groups that are made to spike together within a jitter window of $\pm 1$~ms. The groups are set to spike sequentially with a delay of $\delta = 5$ ms and a repetition period of $\Delta =20.5$ ms.

\subsection{Simulated data from Leaky-Integrate-and-Fire Networks}
We applied our algorithm to more realistic spiking computational datasets obtained by simulating neuronal networks of excitatory and inhibitory Leaky-Integrate-and-Fire (LIF) neurons~\citep{Koch1999}.

\begin{figure*}[tb!]
  \centering
  \includegraphics{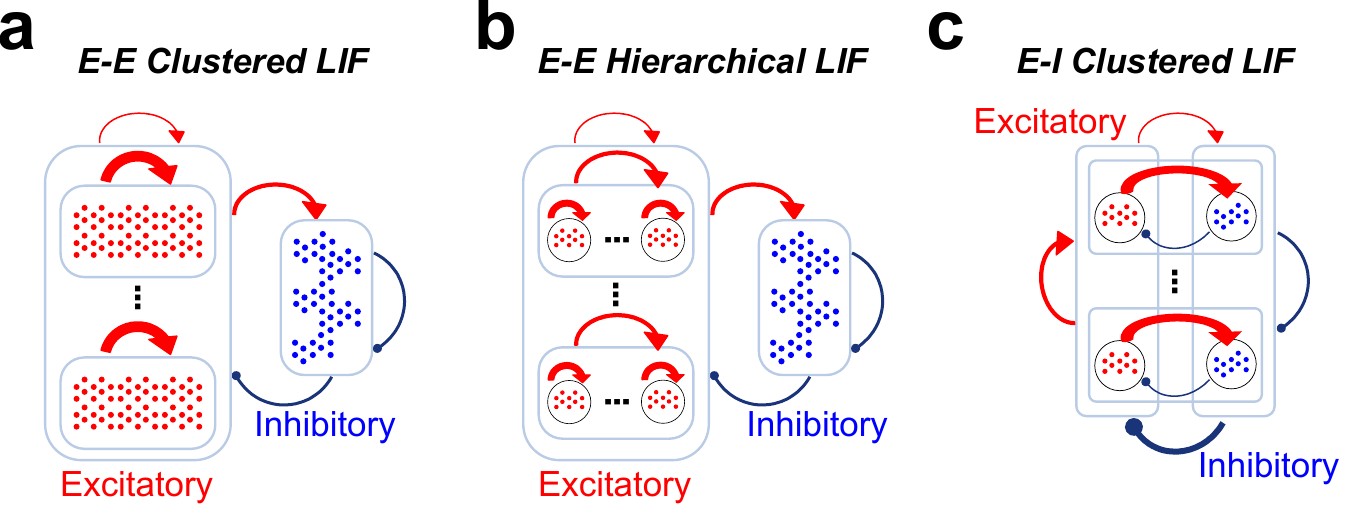}
 \caption{Schematic wiring diagrams of the three LIF networks used in this work: \textbf{(a)}~an E-E clustered LIF network; \textbf{(b)} an E-E hierarchical LIF network; \textbf{(c)} an E-I clustered LIF network.
 Arrow thickness is proportional to the strength of the connection. For the parameters used in our simulations, see Table~\ref{tab:parameters}.}
  \label{fig:3}
\end{figure*}

\begin{table*}[tb!]
\begin{center}
\scriptsize
\begin{tabular}{l ccccc ccl}
\toprule
 & \multicolumn{8}{c}{\textbf{Probabilities}}  \\
\cmidrule(r){2-9}
 & $p^{II}$    & $p^{IE}$ & $p^{EI}$ & $p^{EE}$ & $p^{EE}_\text{sub}$ & $p^{EE}_\text{sub,sub}$ & $p^{EI}_\text{sub}$ & $p^{IE}_\text{sub}$  \\
\midrule
E-E Clustered & $0.5$ & $0.5$ & $0.5$ & $0.167$ & $0.5$ & --- &--- &--- \\
E-E Hierarchical  & $0.5$ & $0.5$ & $0.5$ & $0.15$ & $0.3$ & $0.99$ &--- &---\\
E-I Clustered  & $0.5$ & $0.454$ & $0.526$ & $0.2$ & --- & --- &$0.263$& $0.90$  \\
 \midrule
 & \multicolumn{8}{c}{\textbf{Weights}} \\ \cmidrule(r){2-9}
  & $w^{II}$    & $w^{IE}$ & $w^{EI}$ & $w^{EE}$ & $w^{EE}_\text{sub}$ & $w^{EE}_\text{sub,sub}$ & $w^{EI}_\text{sub}$ & $w^{IE}_\text{sub}$  \\
 \midrule
E-E Clustered  & $-0.04$ & $0.01$ & $-0.025$ & $0.012$ & $0.0144$ & --- &---&---\\
E-E Hierarchical   & $-0.04$ & $0.01$ & $-0.03$ & $0.012$ & $0.012$ &  $0.014$& ---&---\\
E-I Clustered  & $-0.04$ & $0.0086$ & $-0.032$ & $0.0155$ & --- & --- &$-0.0123$& $0.0224$\\
\bottomrule
\end{tabular}
\end{center}
\caption{\rm Parameters for the simulated LIF networks. Connection probabilities ($p^{XY}$) and weights ($w^{XY}$) between different unit types: excitatory (E) and inhibitory (I), \textit{e.g.},  $p^{EI}$ is the connection probability from inhibitory to excitatory units.
For the clustered networks, the average E-E connection probability was kept constant at 0.2. For a schematic representation of the wiring diagrams, see Fig.~\ref{fig:3}.}
\label{tab:parameters}
\end{table*}

\subsubsection{The excitatory and inhibitory LIF units}
The non-dimensionalized membrane potential $V_i(t)$ for neuron $i$ evolved according to:
\begin{equation}
  \dfrac{d V_i(t)}{dt} = \dfrac{\mu_i - V_i(t)}{\tau_m} + I_\text{S},
\end{equation} 
where the constant input term $\mu_i$ was chosen uniformly in the interval $[1.1, 1.2]$ for excitatory neurons and in the interval $[1, 1.05]$ for inhibitory neurons.  
Both excitatory and inhibitory neurons had the same firing threshold of $1$ and reset potential of $0$.
Note that although the input term is supra-threshold, balanced inputs guaranteed that the average membrane potential remained sub-threshold \citep{Litwin-Kumar2012,Vreeswijk1998}.
Membrane time constants for excitatory and inhibitory neurons were $\tau_m = 15$~ms and 
$\tau_m = 10$~ms, respectively, and the refractory period was $5$~ms for both excitatory and inhibitory neurons.
The synaptic input from the network was given as: 
\begin{equation}
I_\text{S} = \sum_{i \leftarrow j}w_{i \leftarrow j}g^{E/I}_j(t),
\end{equation} 
where the $i\leftarrow j$ denotes that there is connection from neuron $j$ to neuron $i$, and $w_{i \leftarrow j}$ denotes the weight of this connection (see next section for the weight settings).
The synaptic inputs $g^{E/I}$ were increased step-wise instantaneously after a presynaptic spike ($g^{E/I} \to g^{E/I} +1$) and then decayed exponentially according to:
\begin{equation}\label{eq:synapse}
    \tau_{E/I} \dfrac{d g^{E/I}}{dt} = - g^{E/I}(t),
\end{equation}
with time constants $\tau_E =3$ ms for an excitatory interaction, and $\tau_I =2$ ms if the presynaptic unit was inhibitory.

\subsubsection{Network Topologies and Weight Matrices}
LIF excitatory and inhibitory units in a proportion of $4:1$ were interconnected with three different network topologies. The resulting networks were simulated with a $0.1$ ms time step. The connection probabilities and weights between the different types of neurons for these three LIF networks are shown in Table~\ref{tab:parameters} and the schematic of the different wiring diagrams is shown in Fig.~\ref{fig:3}.

\paragraph*{Network with clustered excitatory connections (E-E clustered)}
We first constructed a LIF network with clustered excitatory units: each excitatory neuron belongs to a group of units more strongly connected to each other than to units outside the group (Figure \ref{fig:3}a). The network also included unclustered inhibitory units, which ensured that the network was balanced. These networks display temporally-structured spike-train activity~\citep{Litwin-Kumar2012}, and are used here as a test-bed for cell-assembly detection from spiking dynamics.

\paragraph*{Network with hierarchical excitatory connections (E-E hierarchical)}
In a similar fashion, we developed a LIF network where excitatory units belonged to a hierarchy of groups (Figure \ref{fig:3}b).
For this, we split the population of excitatory units into nested clusters, such that each group was sub-divided into smaller groups with increasing internal connectivity. The inhibitory neurons remained unclustered.  

\paragraph*{Network with excitation to inhibitory clustered feedback loops (E-I clustered)}
Finally, we have developed a LIF network to study the dynamical spiking patterns originated by networks in which excitatory and inhibitory neurons are co-clustered, as shown in Figure \ref{fig:3}c.
In this case, whereas the excitatory-to-excitatory and inhibitory-to-inhibitory couplings were kept uniform,  we introduced structural features in the connections between distinct neuron types. 
In particular, each subset of excitatory units was more strongly connected to a subset of inhibitory units. This group of inhibitory units, in turn, had a weaker feedback to its associated excitatory neuron group, as compared to the rest of the graph. 
Every unit was part of one such functional group comprising both excitatory and inhibitory units.

\begin{figure*}[hbt!]
  \centering
  \includegraphics{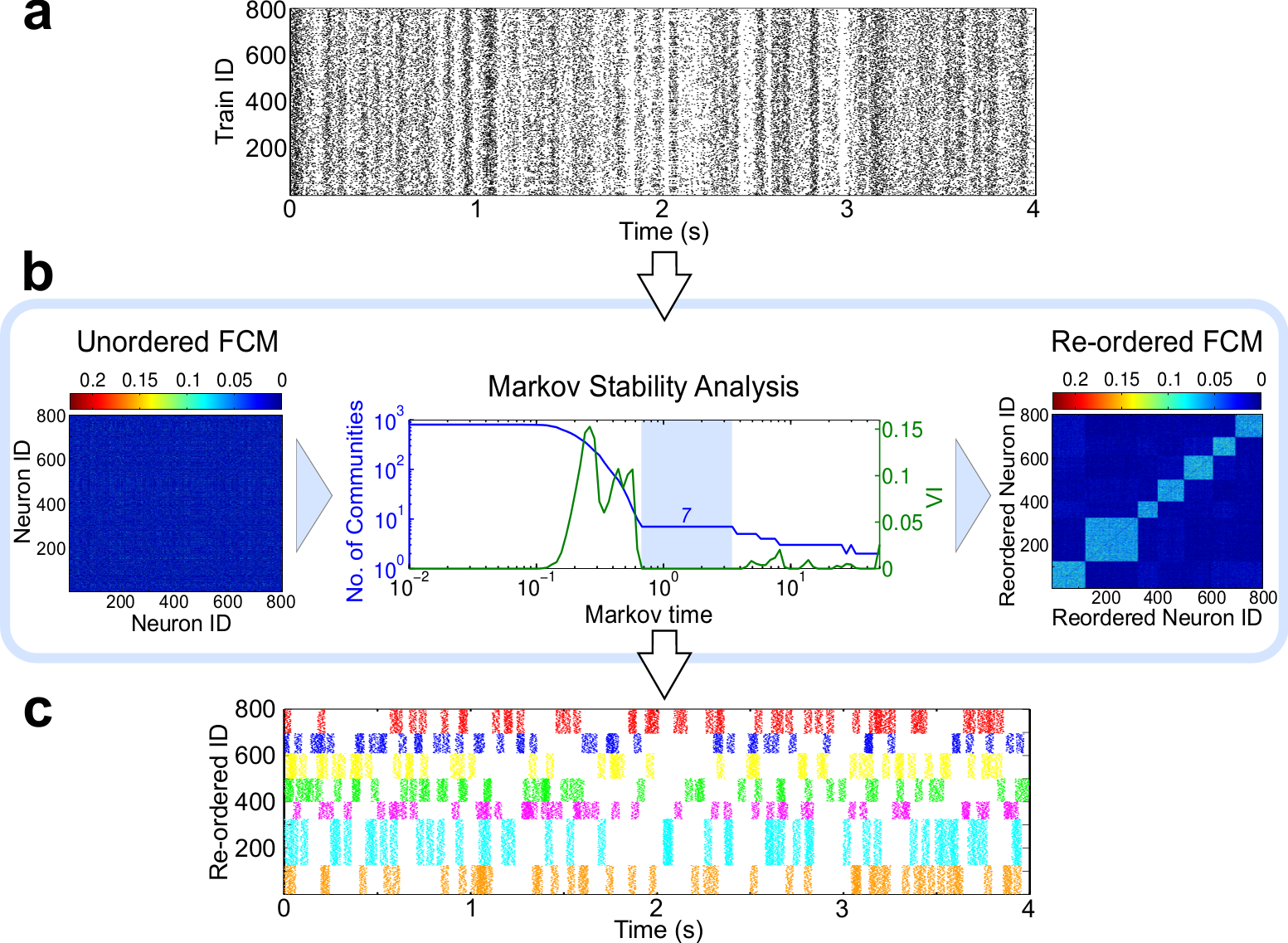}
  \caption{Markov Stability analysis of a synthetic data set.
  \textbf{(a)} Unsorted raster plot of a population of 800 spike-trains obtained from
  7 groups of different sizes. Each `cell assembly' fires at different times with
  varying amounts of jitter.
  \textbf{(b)} FCM from the unsorted spike train rastergram, followed by the Markov Stability plot and the FCM reordered according to the partition into 7 groups obtained by the algorithm. Note the long plateau (blue shaded) around Markov time $t_M=1$ with $VI=0$, indicating the presence of a robust partition with 7 groups.
  At later Markov times, the algorithm detects other robust coarser partitions corresponding to
  aggregates of the seven groups with similar firing patterns.
  \textbf{(c)} Color-coded raster plot reordered according to the partition obtained in
  \textbf{(b)}.   }
  \label{fig:4}
\end{figure*} 

\subsection{Experimental data}

\subsubsection{Retinal Ganglion Cell recordings from mouse and salamander}
These datasets were kindly provided by the lab of Markus Meister.
Multielectrode recordings were performed as described previously \citep{Meister1994}, following protocols approved by the Institutional Animal Care and Use Committee at Harvard University and at the California Institute of Technology.
Dark-adapted retina isolated from a larval tiger salamander (\textit{Ambystoma tigrinum}) or adult mouse (\textit{Mus musculus}; C57BL/6) was placed on a flat array of 61 extracellular electrodes with the ganglion cell side down.
The salamander retina was superfused with oxygenated Ringer's medium (in mM: NaCl, 110; NaHCO$_3$, 22; KCl, 2.5; MgCl$_2$, 1.6; CaCl$_2$, 1; and D-glucose, 10; equilibrated with 95\% O$_2$ and 5\% CO$_2$ gas) at room temperature. The mouse retina was perfused with oxygenated Ame's medium (Sigma-Aldrich; A1420) at 37$^\circ$C.

Recordings were made with a custom-made amplifier and sampled at 10 kHz.
Spike sorting was performed offline by analyzing the shape of action potentials on different electrodes~\citep{Pouzat2002,Gollisch2008}.
The spike-triggered averages (STAs) and receptive fields of the salamander retinal ganglion cells (RGCs) were determined by reverse correlation to a checkerboard stimulus flickering with intensities drawn from a normal distribution.
Singular-value decomposition of the spatio-temporal receptive field allowed the extraction of the temporal filter of every RGC receptive field \citep{Gollisch2008}.

\subsubsection{Hippocampal CA1 and CA3 recordings from rats under a spatio-temporal task}
We analyzed spike trains obtained by \citet{Diba2007} from hippocampal neurons of rats moving along a linear track implanted with silicon probe electrodes along CA1 and CA3 pyramidal cell layers in left dorsal hippocampus.

\subsection{Performance of the method and comparisons to other techniques}
In those examples where the results could be compared against a ground truth, the performance of the method was determined by the percentage of correctly classified neurons (hit rate) relative to the true membership in the data.

We have compared the performance of our methodology with two other popular community detection techniques:
Modularity optimization (two variants) using the code provided and explained in \cite{Humphries2011};
and standard agglomerative hierarchical clustering using the nearest distance linkage criterion as implemented in MATLAB. 

\section{Results}

\subsection{Assessing the algorithm with synthetic datasets}

We first tested our method on synthetic spike-train datasets to evaluate its performance and to showcase its distinct capabilities compared to other methodologies.

\subsubsection{Analysis of synthetic data with embedded cell-assemblies}
As a first illustration, Figure~\ref{fig:4} shows the application of our method to a synthetic spike dataset with inherent group structure (see Materials and Methods).
A population of 800 units was divided into 7 differently sized groups comprising 75 to 200 units.
The average spiking frequency for all groups was 12 Hz with $\pm20$~ms jitter around the uniformly chosen firing times within the total length of 4 s. 

\begin{figure}[htb!]
  \centering
  \includegraphics{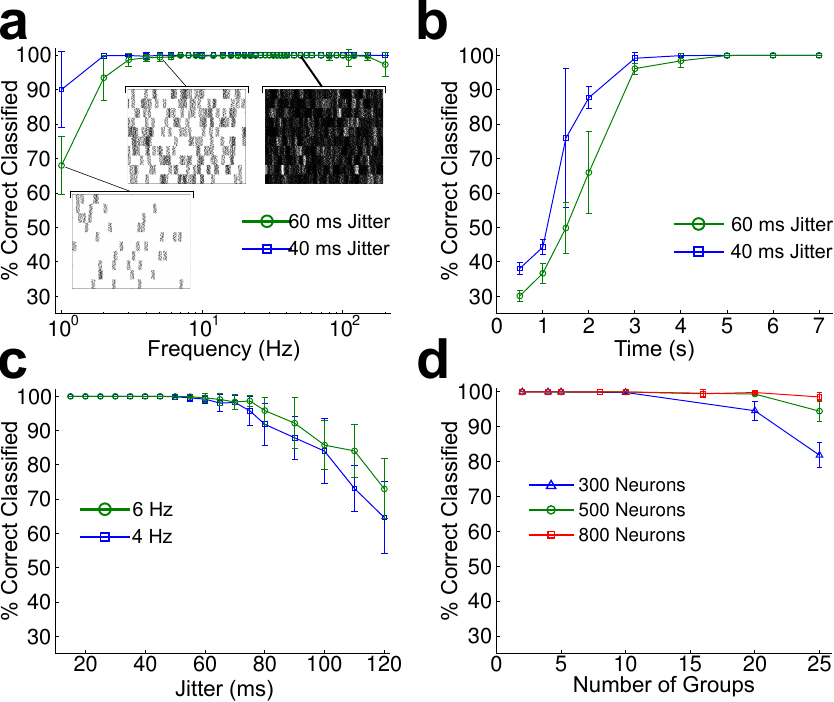}
  \caption{Assessing the performance of the clustering algorithm using synthetic data.
  \textbf{(a)}~At very low firing frequencies, the classification performance is low due to a small number of spikes per neuron. Performance quickly improves with increasing firing frequency.
  \textbf{(b)}~The classification performance improves as the duration of the recording increases.
   \textbf{(c)}~As the jitter increases, the classification performance degrades.
  \textbf{(d)}~The classification performance degrades mildly as the number of groups to be detected increases.}
  \label{fig:5}
\end{figure}

\begin{figure*}[thb!]
  \centering
  \includegraphics{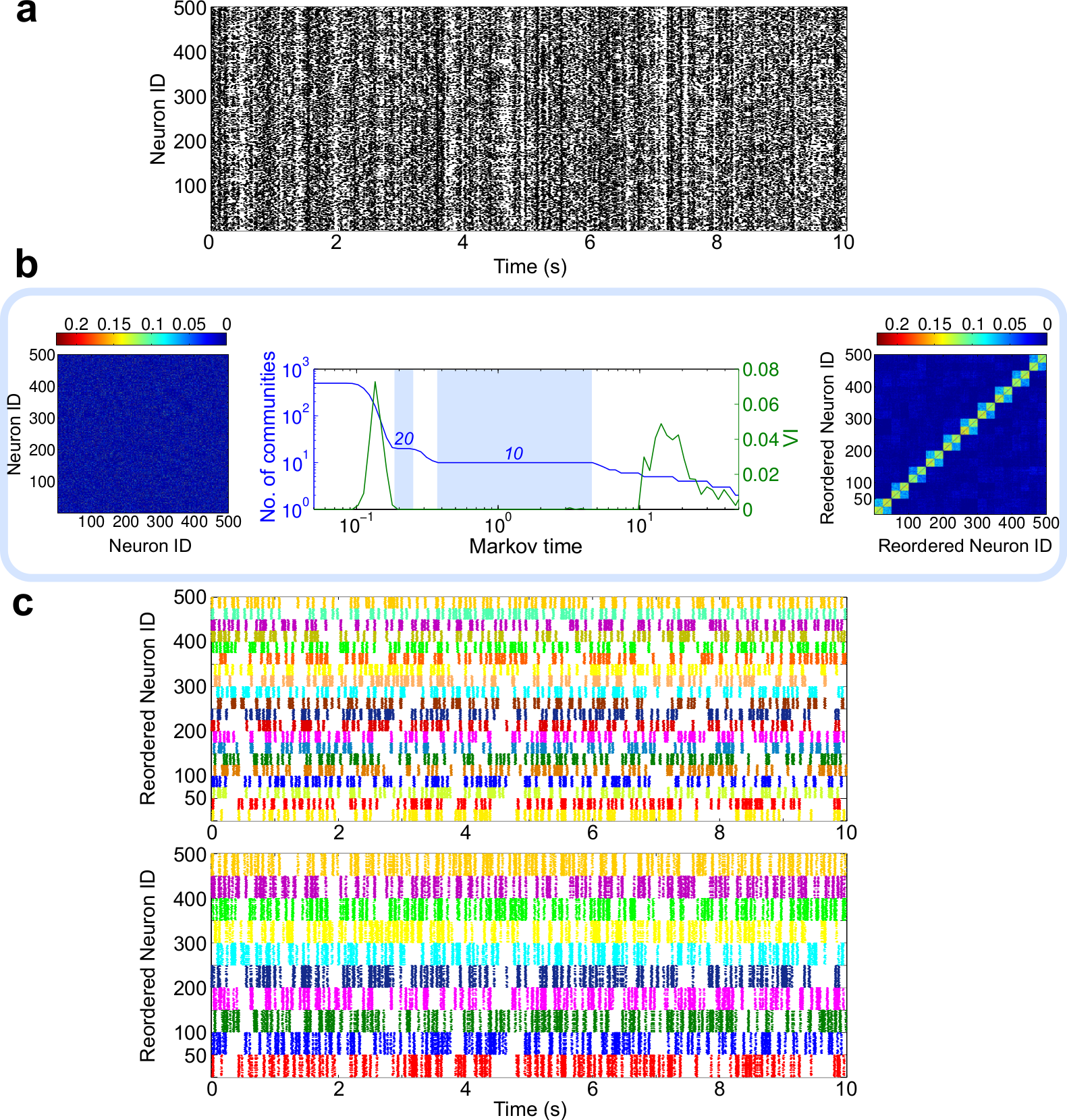}
 \caption{Detecting hierarchically structured spike train communities in synthetically generated data.
  Synthetic data of 500 units clustered into 10 groups with 2 subgroups each (20 subgroups in total).
  \textbf{(a)} Unsorted raster plot of data.
  \textbf{(b)} Markov Stability analysis of the associated FCM.
  Clear plateaus indicate the presence of robust partitions into 20 and 10 communities, with
  classification accuracy of 100\% in both cases.
  \textbf{(c)} Sorted raster plots for the finer (20 groups, top panel) and coarser (10 groups, bottom panel) partitions revealing the hierarchical organization in the data.}
  \label{fig:6}
\end{figure*}

Figure~\ref{fig:4}a displays the raster data prior to clustering. Figure~\ref{fig:4}b shows the functional connectivity matrix (FCM) calculated from the spike trains used in  the Markov Stability analysis leading to the identification of a robust seven-community partition leading to a reordered FCM.
The raster plot reordered according to the communities identified by our algorithm is shown in Figure~\ref{fig:4}c.
The detected partition into 7 groups corresponds to an extended plateau in Markov time (from $t_M=0.68$ to $t_M=3.47$) with $VI=0$, in which all the neurons were correctly clustered.
Note that the algorithm detects other partitions with relatively long plateaux in Markov time, although their variation of information is non-zero.
In particular, a relatively robust partition into 3 clusters between $t_M=8.21$ and $t_M=25.12$ is detected corresponding to a coarser grouping of the seven groups embedded in our data.  

\begin{figure*}[htb!]
  \centering
  \includegraphics{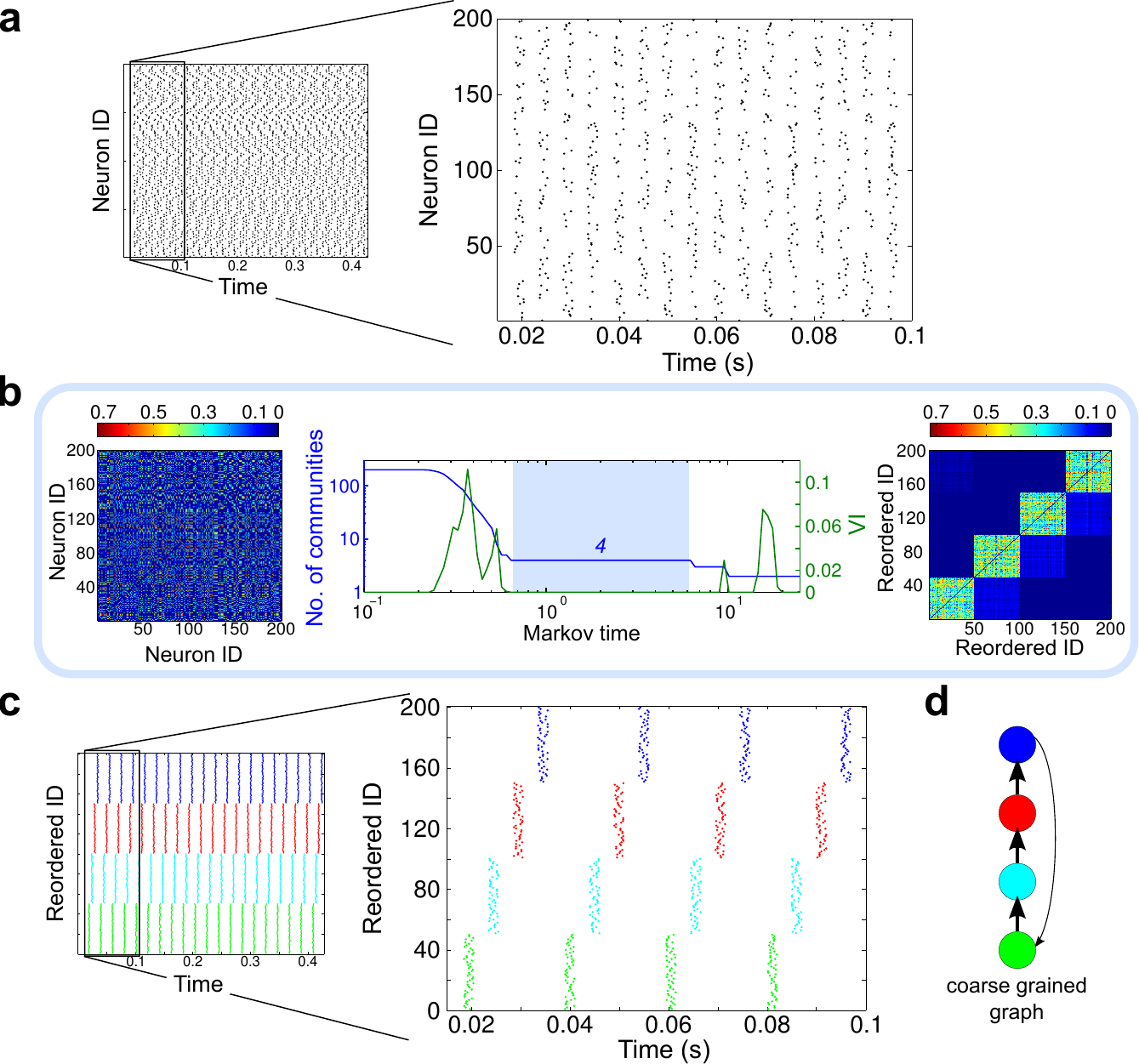}
 \caption{Analysis of feedforward-like firing patterns.
  \textbf{(a)} Unsorted raster plot of the synthetic data and zoom-in.
  \textbf{(b)} Markov Stability analysis of the FCM identifies a robust partition into 4 groups, with
  100\% classification accuracy. Note that the FCM is asymmetric, thus revealing the directionality of the data.
   \textbf{(c)} Color-coded raster plot and zoom-in color-coded according to the partition found reveals the feed-forward functional relationship in the data.
   \textbf{(d)} Coarse-grained representation of the functional connectivity network found from the clustering. For a second example, see supplementary information.}
   \label{fig:7}
\end{figure*}

To assess the performance and robustness of the procedure, we determined the percentage of correctly classified neurons under a variety of noise conditions, different amounts of data, and other sources of variability. 
Figures \ref{fig:5}a,b demonstrate the accuracy of classification for 500 units with 10 cell assemblies when the number of observed spiking events is varied.
In Figure \ref{fig:5}a, spiking datasets of fixed length 4~s are analyzed as the firing frequency is increased.
As expected, the performance degrades at very low and very high spike frequencies when the number of firing events is either too low or too large to distinguish the groupings (see insets). 
 This effect can be reduced with increasing spike-train lengths as shown in Figure \ref{fig:5}b where the performance improves as we increase the duration of the recording for a fixed firing rate (4 Hz). For short recordings, spurious correlations in the firing events degrade the performance, which consistently improves as the duration increases reaching 100\% accuracy for recordings of length above~4~s.
To assess the effect of jitter, we checked that the above examples (Figure \ref{fig:5}a-b)  show similar behavior for different amounts of jitter ($\pm 40$~ms, $\pm 60$~ms).
The performance degrades only when the jitter is increased strongly (Figure \ref{fig:5}c).
Finally, we assessed the sensitivity of the method and its ability to detect an increasing number of groups in a population of given size (Figure \ref{fig:5}d).
As expected, the accuracy of the classification drops, but only mildly, as we increase the number of groups to be detected. Note that this is due partly to an entropic effect: a correct assignment among a larger number of groups conveys more information than a correct decision between fewer groups.
Hence, the decrease in performance is even less dramatic if corrected for this effect.

\subsubsection{Analysis of synthetic hierarchical spiking patterns}
Hierarchical neuronal connectivity~\citep{McGinley2013,Savic2000,Ambrosingerson1990} can lead to spiking dynamics with temporal structure at different scales.
One advantage of using Markov Stability is its ability to detect hierarchical structure in data without \textit{a priori} knowledge of such relations.
To showcase this capability, we created synthetic data sets with embedded hierarchical relationships (see Materials and Methods).
Figure~\ref{fig:6} illustrates the analysis of the spiking dynamics from 500 units, which are split into 10 groups of co-firing units with each group further sub-divided into two subgroups that fire together more frequently. This results in a hierarchical organization of $20 \to 10$  subgroups.
Using the same methodology as above, our analysis reveals two extended plateaus with $VI=0$, for 20 and 10 groups (Figure \ref{fig:6}b).
The sorted raster plots for the 20 and 10 groups, shown in Figures \ref{fig:6}c, correspond to 100\% correct classification.
As we will demonstrate below in the context of LIF networks, this consistent multi-scale detection of cell assemblies is a distinct feature of our methodology, which is not present in many other methods which only detect groupings at a particular level of granularity \citep{Fortunato2010}.

\subsubsection{Analysis of synthetic feedforward spiking patterns}
To highlight the capability of our framework to deal with directed dynamical patterns, we show how feedforward-like functional patterns in the data lead to a pronouncedly asymmetric FCM, which can then be analyzed with Markov Stability.
Synthetic spiking patterns were generated in which 4 groups of 50 neurons (with jitter) spiked 20 times emulating synchronous activity in feedforward networks (see Materials and Methods).
As shown in Figure~\ref{fig:7}, our method is able to detect feedforward patterns between cell assemblies: the corresponding Markov Stability plot shows a robust and extended plateau with 4 communities with 100\% classification accuracy revealing an effective coarse-grained description of a functional feedforward network.

This is an instance in which the directed nature of our FCM, together with the fact that Markov Stability can detect communities in directed networks, leads to the detection of cell assemblies with directed, causal relationships. 
Indeed, there are instances of directed functional couplings~\citep{Rosvall2008} in which using symmetric measures will lead to different cell assemblies to those obtained if directionality is taken into account.
Hence for some networks, directionality is absolutely essential for proper clustering (see supplementary information).
Within our framework, the importance of directionality can be tested by including or disregarding directionality in the analysis and comparing the outcomes.

\subsection{Detecting cell assemblies in simulated dynamics of Leaky-Integrate-and-Fire (LIF) networks}
Beyond purely synthetic datasets, we now consider three examples of simulated dynamics of LIF networks, which exhibit a range of features of relevance in realistic neural networks. LIF networks provide a simple, controlled testbed to assess our framework on network dynamics broadly used in computational systems neuroscience.

\subsubsection{Cell assemblies in LIF networks with clustered excitatory connections}
It was demonstrated recently that balanced LIF networks with clustered excitatory connections can display network dynamics in which the clustered neurons spike in a coordinated manner over long timescales \citep{Litwin-Kumar2012}.
We implemented such a LIF-network to determine if our framework was able to recover the underlying \textit{structural} connectivity directly from the observed spiking dynamics.

\begin{figure}[tb!]
  \centering
  \includegraphics{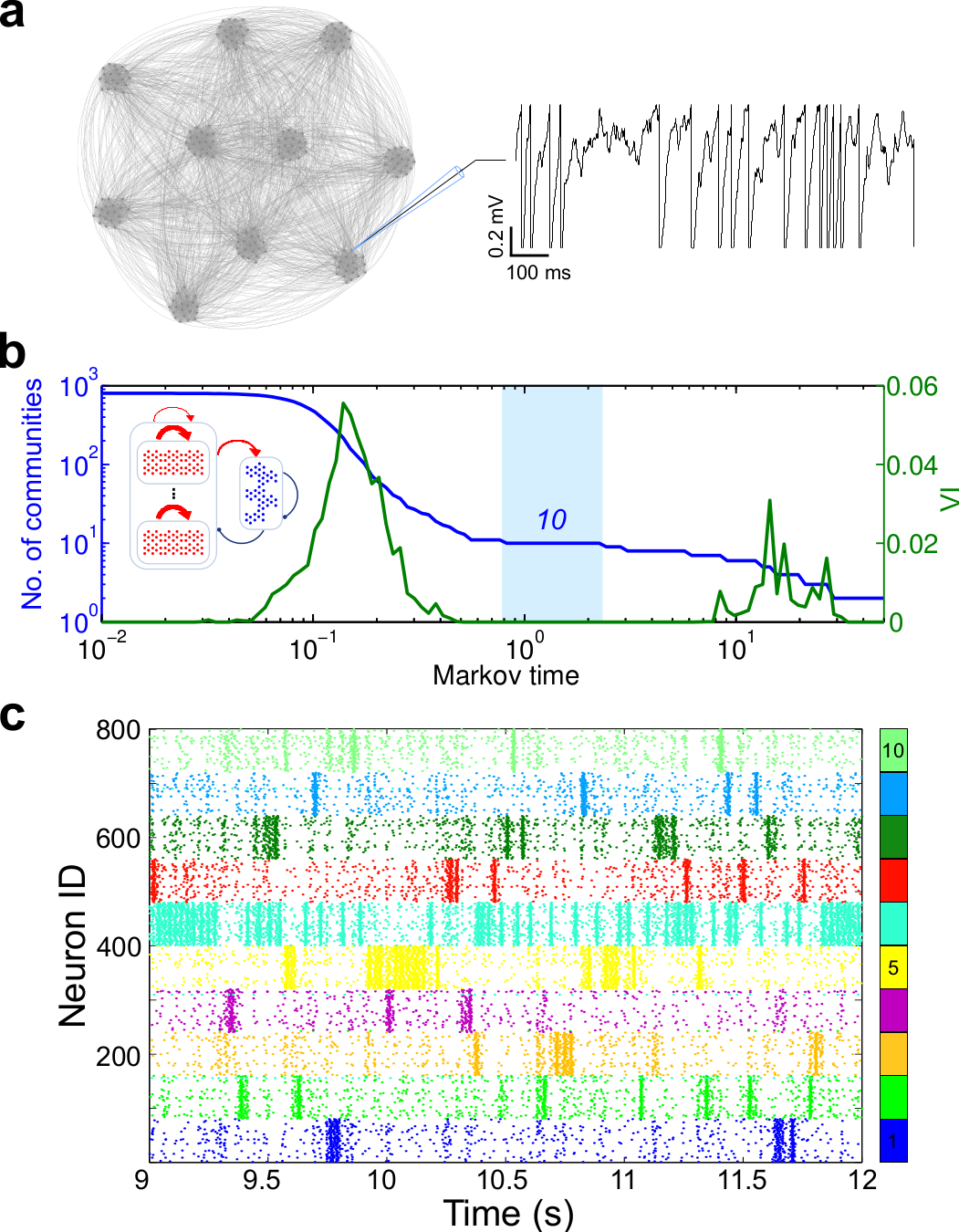}
 \caption{Detecting cell assemblies in spiking data from E-E clustered LIF networks.
  \textbf{a} Schematic of the excitatory connectivity of the LIF network. The 800 excitatory units were split into 10 groups such that the intra-group connection probability and synaptic strength were larger than the inter-group values. The network was balanced with 200 unclustered inhibitory units. An example of the simulated membrane potential traces for an excitatory unit is also shown. 
  \textbf{b} Markov Stability analysis of the corresponding FCM. There is a clear plateau with $VI=0$ for a split into 10 groups (blue shaded).   Inset: Schematic of network topology.
  \textbf{c} Color coded raster plot according to the partition obtained. Units are ordered consecutively according to their grouping in the underlying LIF topology. The correct grouping is also indicated by the colored band on the side. These cell assemblies exhibit clear bands of activity. Only 5 neurons were misclassified relative to the imposed structure (99.5\%).}
  \label{fig:8}
\end{figure}

Figure \ref{fig:8}a depicts a schematic of the structural connectivity imposed on the simulated E-E clustered network: 800 excitatory units were split into 10 groups such that the connection probability and synaptic strengths within each group were larger than the inter-group values (see Materials and Methods).
The network was balanced by 200 inhibitory units that were uniformly connected (\textit{i.e.,} unstructured). The spiking dynamics of the excitatory units were then analyzed using our framework:
the spike trains of the 800 excitatory units were used to generate the FCM, and Markov Stability revealed a clear plateau with $VI=0$ corresponding to a partition into 10 groups with a
classification performance of 99.5\% with respect to the embedded structural groups. 

The results of our method contrast with other commonly used methods. For instance, applying hierarchical clustering to the FCM achieves only a classification performance of at best 22\%, and does not provide a clear criterion for determining the number of groups present.
Similarly, we apply Modularity optimization using the two versions implemented by \cite{Humphries2011} in conjunction with our FCM matrix. In this case, Modularity imposes an intrinsic scale leading to the identification of 8-13 groups, with a classification performance of 49-68\% depending on which of the two versions of the optimization is used.
Let us remark that we also applied hierarchical clustering directly to the time-series.
However, as it leads to similarly poor performance, we do not report these results here.

As explained above, an advantage of Markov Stability is that it does not impose \textit{a priori} the scale or the number of clusters to be detected. Instead, the method scans across all scales and extracts robust, meaningful partitions at different levels of granularity, thus revealing potentially relevant partitions in the data. In clear-cut cases, such as the simple synthetic datasets studied above, the method reveals unequivocally the partitions embedded in the data. In general, however,  and especially for noisy data, it is not expected that a unique partition is found. Rather, a set of candidate partitions will emerge. This is observed in the analysis of this LIF network. Figure \ref{fig:8}b shows that the partition into 10 communities is the clearest choice for this dataset (
longest plateau in Markov time with $VI=0$). However, other good candidate partitions include: one into 11 communities, which is similar to the grouping into 10 albeit with an additional split for a small group of neurons, and a partition into 8 communities obtained by the merging of 4 of the groups into 2 groups.
Our methodology provides candidate partitions at different levels of resolution based on their robustness (i.e., long plateaux in Markov time, drops in VI) as a guide for the analysis, which can then be complemented with further biological knowledge.

\begin{figure*}[htb!]
  \centering
  \includegraphics{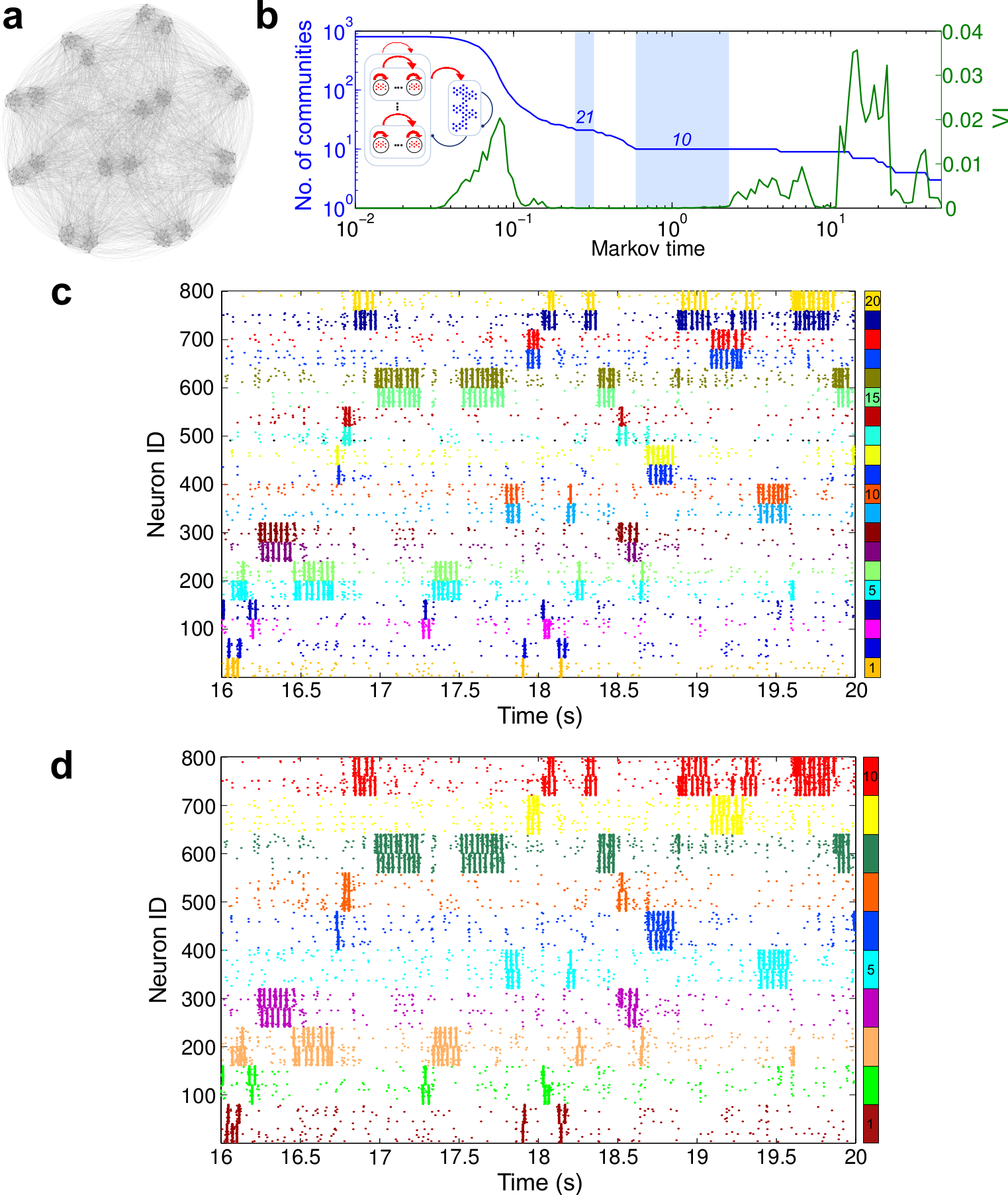}
\caption{Detecting cell assemblies in spiking data from E-E hierarchically clustered LIF networks.
  \textbf{(a)} Schematic of the network with excitatory units split into 10 groups which were further sub-divided in 2 subgroups each.
  \textbf{(b)} Markov Stability plot of the analysis reveals two robust partitions with 21 and 10 communities.  Inset: Schematic of network topology.
  \textbf{(c)} Color coded raster plot of the clustering into 21 communities with classification accuracy of 99.9\% (one neuron was misclassified). Units are ordered consecutively according to their grouping in the underlying LIF topology. The correct grouping is also indicated by the colored band on the side.
   \textbf{(d)}  Color coded raster plot of the 10 community clustering with an accuracy 100\%.}
  \label{fig:9}
\end{figure*}

\subsubsection{Cell assemblies at multiple levels of granularity in hierarchical LIF networks}
To further test the multiscale capabilities of our method, we evaluated how the algorithm would perform on the analysis of a LIF network with a hierarchical structure. The population of 800 excitatory units was divided into 10 groups with 2 subgroups each, \textit{i.e.,} 20 groups in two levels of a hierarchy (Figure~\ref{fig:9}a). 
As before, the network was balanced with 200 unclustered inhibitory units. 
The Markov Stability analysis displayed in Figure \ref{fig:9} reveals two robust partitions: one into 21 communities and one into 10 communities with classification accuracies of 99.9\% and 100\%, respectively, and which correspond to the two levels of resolution.
Our method also finds other suitable but less robust candidate partitions, \textit{e.g.}, one into 9 communities obtained by merging 2 of the top-level communities into a single group, or one into 4 groups obtained by similar mergers.  

As stated above, other commonly used methods are unable to detect these multiple levels of granularity.
For instance, Modularity (using again the implementation of \cite{Humphries2011}) finds a partition at one particular scale (10 communities, 41-80\% accuracy for the two versions)
and cannot detect the presence of the finer grouping.
The application of hierarchical clustering, 
leads to an agglomerative tree with no better accuracy than 41\% at any level of granularity and no clear criterion to detect the number of communities present.

\subsubsection{Mixed cell assemblies with excitatory and inhibitory units in LIF networks}
Hitherto we have only considered the clustering of excitatory units.
However, functional groups of neurons may be composed of mixtures of excitatory and inhibitory neurons~\citep{Buzsaki2010}.
The definition of our spike-train similarity allows for the detection of such relationships by incorporating the biophysical effect of both excitatory and inhibitory neurons on their postsynaptic neurons (\textit{i.e.,} EPSPs \textit{vs.} IPSPs;  see Figure \ref{fig:1}A). To determine how our method would perform in a context where mixed functional groups are present, we created a LIF network with an embedded structure between excitatory and inhibitory units. As shown in Figure \ref{fig:6}A,  the coupling between alike neuron types is uniform but we create preferential coupling between subsets of excitatory and inhibitory neurons, \textit{i.e.}, each subset of excitatory units is preferentially connected to a subset of inhibitory units (relative to all other inhibitory ones) and, in turn, this subset of inhibitory units feeds back weakly to their corresponding subset of excitatory units (relative to all other excitatory units).
Our simulated LIF network included 10 such groups with 80 excitatory neurons and 20 inhibitory neurons per group for a total of 1000 neurons.

\begin{figure}[htb!]
  \centering
   \includegraphics{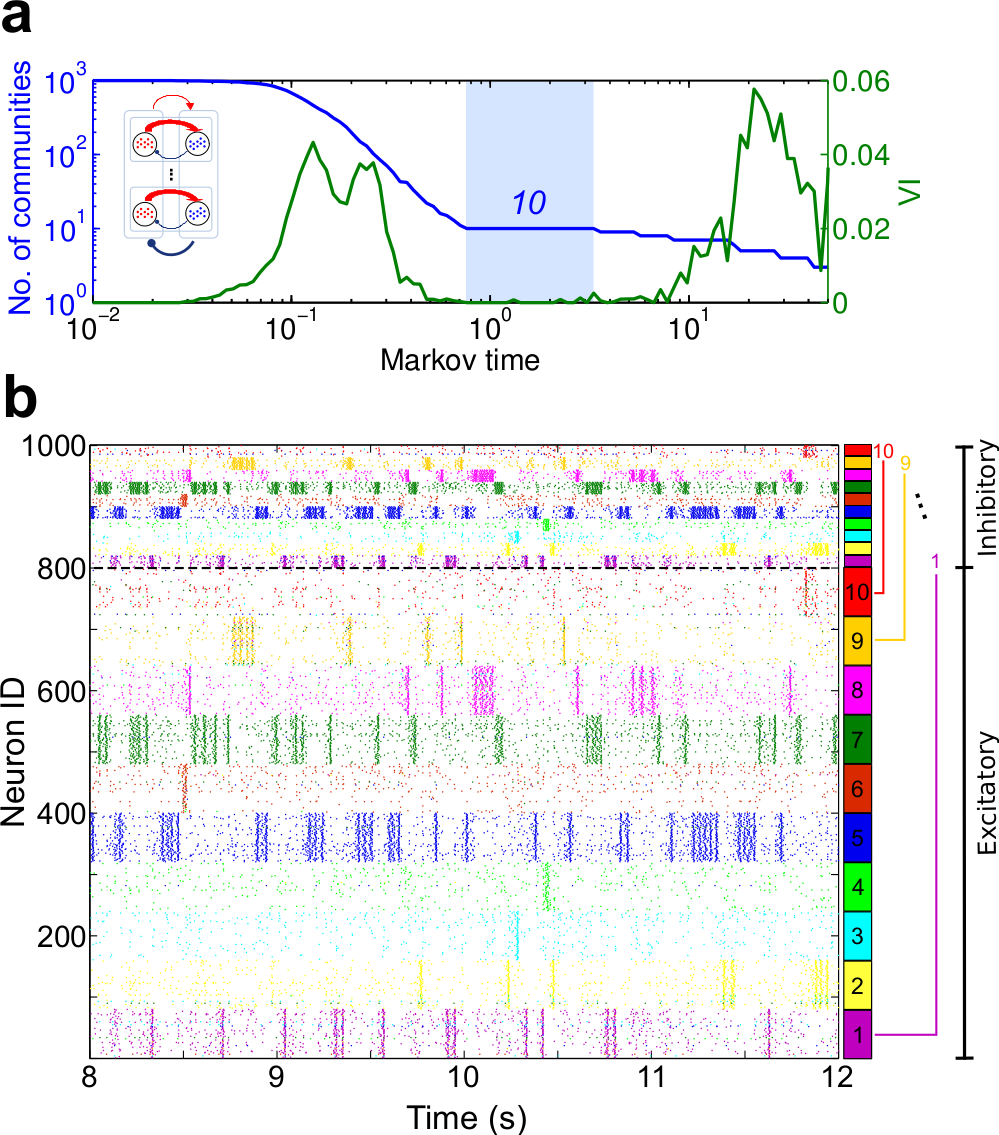}
 \caption{Detecting cell assemblies in spiking data from E-I clusterd LIF networks. The network has functional groups comprising excitatory and inhibitory neurons.
  \textbf{(a)}~Stability plot of the clustering analysis. Note the clear plateau for 10 communities. Inset: Schematic of network topology.
  \textbf{(b)}~Color-coded raster plot according to the obtained partition.
Note that each group contains excitatory and inhibitory units, as indicated by the color-coded band on the right side which displays the true structural grouping.
For a 20~s simulation, the classification rate was 91.4\%.}
  \label{fig:10}
\end{figure}

The analysis of the dynamics of this LIF network is presented in Figure~\ref{fig:10}.
As indicated by our color coding, we find a robust partition into 10 communities that comprise a combination of both neuron types with a 91.4\% of correctly classified cells according to the embedded structure. Having the ability to account for the role of the inhibitory neurons within a cell assembly may provide a key difference to find a meaningful interpretation of the data.
In our LIF network simulations, we observed that not accounting for cell type differences may result in a drop of up to 20\% in classification performance (data not shown).

\subsection{Applying the algorithm to experimental data}

\subsubsection{Detecting distinct Retinal Ganglion Cells in mouse data}

\begin{figure}[tb!]
  \centering
  \includegraphics{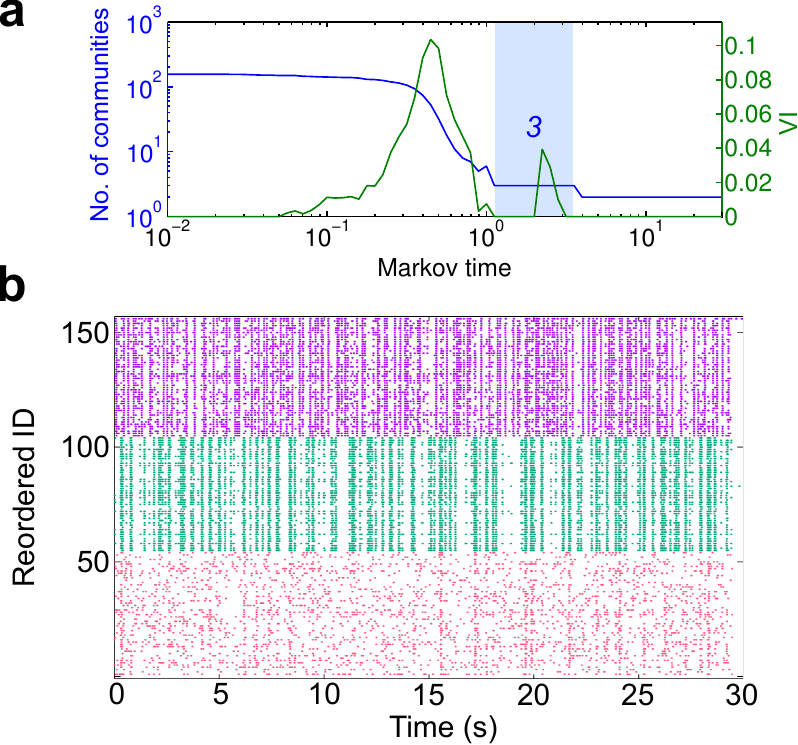}
 \caption{Detecting cells in a  set of extracellular recordings from three mouse retinal ganglion cells (RGCs) to a full field stimulus (52 repetitions for 156 spike trains in total). 
  \textbf{(a)} The Markov Stability plot obtained from all the spike trains reveals the presence of a robust partition into three groups.
  \textbf{(b)} Raster data color coded and reordered according to the communities found.
  The classification was 98.7\% accurate.}
  \label{fig:11}
\end{figure}

As a simple first check of our framework when applied to experimental data, we tested that our algorithm could detect distinct mouse retinal ganglion cells (RGCs) from spike-train data.
Extracellular recordings of a flattened mouse retina were performed while a full field black and white flicker stimulus was repeated 52 times for approximately 28~s (see Materials and Methods).
Data was collected from three different cells that were reliably identified and spiked consistently for every stimulus repetition. 
Spike-triggered averages (STAs) and spike-triggered covariances (STCs) \citep{Schwartz2006} of the three cells were used to characterize the neurons as an ON-cell, an OFF-cell, and a noisy OFF-cell (i.e., an OFF-cell with high trial-to-trial variability).

To test our algorithm, the 52 repetitions from the three neurons were compiled and shuffled randomly into a composite raster plot. The algorithm was then applied to this raster plot so as to find relevant groupings in the 156 spike-trains. 
Our algorithm (Fig.~\ref{fig:11}) reveals a robust partition into three groups corresponding to spike-trains from each cell (98.7\% correctly identified with their original cell). Interestingly, a further plateau at longer Markov times can be seen in Fig. 11a, corresponding to a 2-way partition, in which the two OFF cells are grouped together and the ON cell is separate. The performance accuracy is still 98.7 \%.

Although used here as a check of our method on real data, this analysis also illustrates how our method is able to extract valuable information from the data at multiple levels of resolution directly from the spike-trains.
In contrast, Modularity finds here only a partition into two groups with 60.3-98.1\% correctly identified, a classification performance similar to hierarchical clustering (67\%).

\subsubsection{Detecting classes of Retinal Ganglion Cells in salamander data}
\begin{figure}[tb!]
  \centering
  \includegraphics{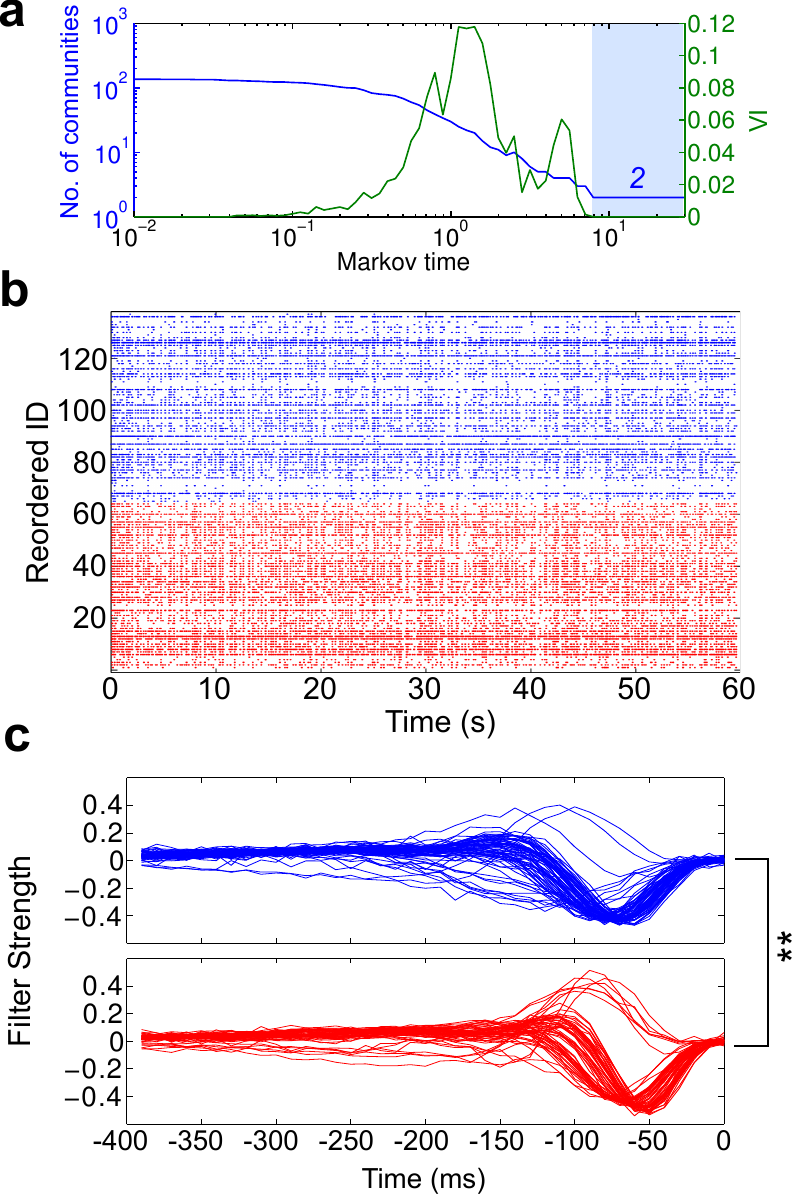}
  \caption{Detecting cell assemblies in spike train recordings from salamander retina RGCs (141 neurons simultaneously recorded with extracellular electrodes).
  \textbf{(a)} The Markov Stability plot obtained from the recordings reveals the presence of a robust partition into two groups.
  \textbf{(b)} Raster data color coded and reordered according to the communities obtained.
  \textbf{(c)} Spike triggered average (STA) responses from all neurons recorded. 
 Each line is an STA for a different cell. The two panels correspond to the two communities (colored as in  \textbf{(b)}) and correspond to transient and sustained RGCs. The full width half maximum of the sustained RGCs (upper panel) is $112.2 \pm 9.1$ ms while for the transient RGCs (lower panel) is  $83.0 \pm 6.4$ ms (mean $\pm$ sem). The difference between the temporal characteristics of the two panels is statistically significant. ($p < 0.01$; see text for details).}
\label{fig:12}
\end{figure}

Next we analyzed a dataset of extracellular simultaneous recordings from multiple RGCs from the salamander retina of three different animals exposed to the same stimulus.
The applied stimulus (i.e., random flickering bars on a screen) entailed both time and spatial components. Approximately 50 neurons were recorded simultaneously from each animal for a total of 141 neurons which were combined into a single raster plot.

Upon application of our algorithm to the spike-train data (Figure~\ref{fig:12}), a robust partition into two groups was observed.
To check if this grouping was meaningful, we studied \textit{ a posteriori}  the STAs of the recorded neurons, which characterized them as mixed groups of ON and OFF cells, yet with different temporal characteristics.
Therefore the two communities found do not correspond to a pure separation into ON and OFF cells but rather to transient and sustained RGC populations responding to fluctuating light intensities on different time-scales~\citep{Awatramani2000}.
Independent statistical confirmation was obtained by checking that the distribution of the full-width-half-maxima of both populations was significantly different between groups (Kolmogorov-Smirnov test to check normality, Wilcoxon rank sum test with $p < 0.01$).

\subsection{Detecting Hippocampal Place Cells in rat recordings}
Finally, the algorithm was applied to CA1 and CA3 hippocampal recordings from a rat moving in a linear track for a water reward~\citep{Diba2007}.
The analyzed data contains 165 neurons which were recorded simultaneously during a series of translocations in which the rat was always traveling in one direction.
The spike trains of the twenty translocations were then spliced together into a raster plot.

\begin{figure*}[tb!]
  \centering
  \includegraphics{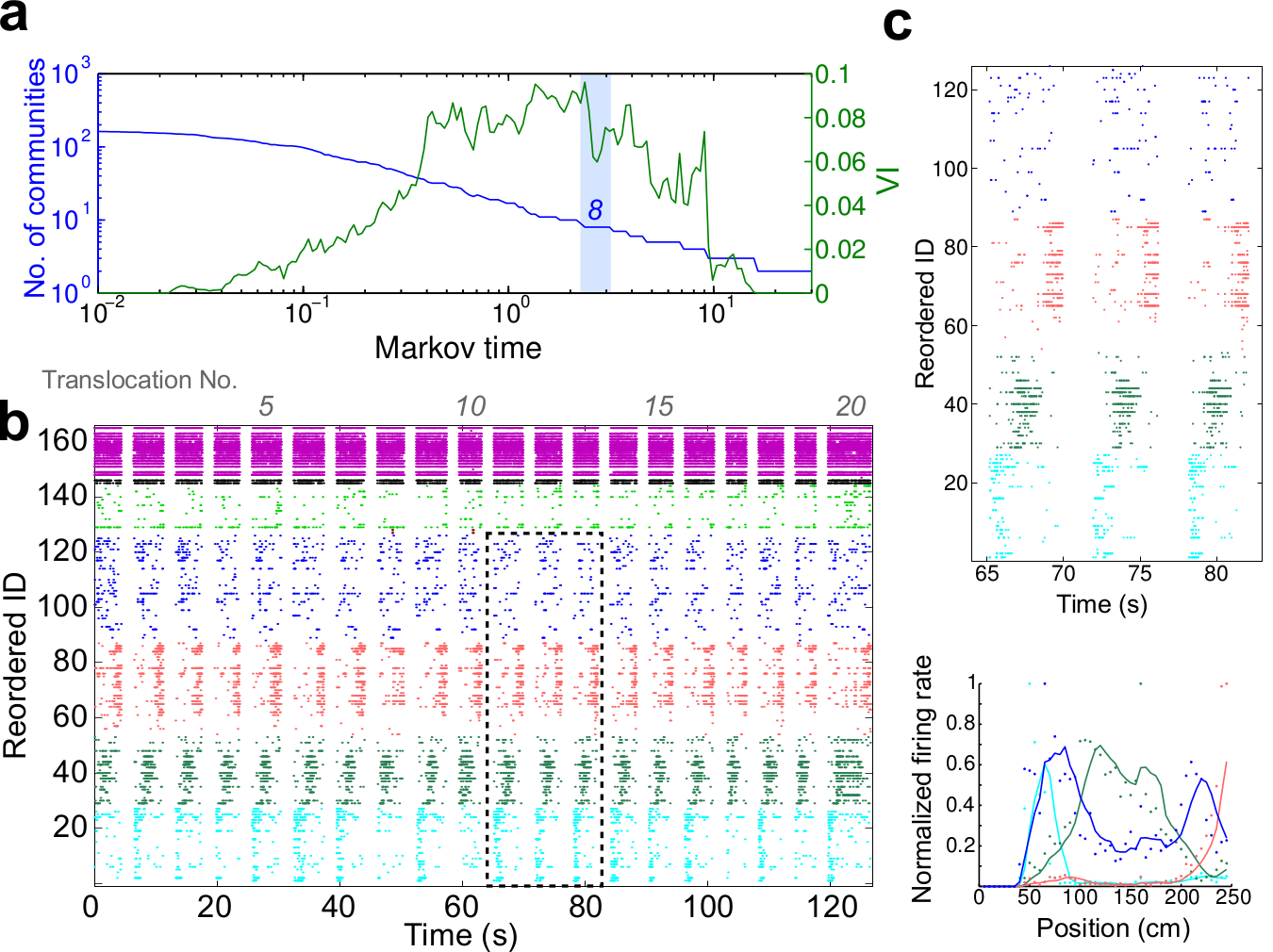}
 \caption{Detection of cell assemblies in recordings from rat hippocampal neurons transversing along a linear track.
  \textbf{(a)}~The Markov Stability plot from the spike-trains shows a small plateau with dip in $VI$ corresponding to a partition into 8 communities.
  \textbf{(b)}~The raster data clustered according to the 8 communities reveals two groups containing the inhibitory neurons (purple and black), and 4 communities with structured time firings (cyan, dark green, red, blue) containing putative place field neuronal groups. These four communities contained 100.0\% of the place cells identified by an independent identification method.
  The columnar gaps separate different translocation events. 
 \textbf{(c)}~Blow-up of the raster plot corresponding to the four place cell communities (top panel) and normalized firing rate of these four groups as a function of distance along the linear track.  Each group favors firing at specific positions along the track.}
  \label{fig:13}
\end{figure*}

Figure~\ref{fig:13} presents the results of our analysis. The FCMs calculated for each translocation were averaged to obtain the FCM. The Markov Stability analysis finds a stable bi-partition, yet 
one of the groups comprised only two neurons of no apparent biological relevance, and this partition was 
not considered further.

An additional long plateau was marked by a dip of $VI$ at $t_M = 2.7$ corresponding to a partition into 8 communities.
We found that two of those communities contain all 18 inhibitory neurons (as identified via spike-sorting in \citep{Diba2007}): the purple group (16 inhibitory neurons) and the black group with the remaining two.
In addition, four of the other communities exhibit structured firing at particular times for every translocation: the first group (cyan) is active at the start of the translocation followed by the dark green group and the red group, while the dark blue group, though less salient, corresponds to cells with firings in between those groups. 
Such groups with localized firing patterns are good candidates to include place cells so we checked \textit{a posteriori} the normalized firing rate of these communities as a function of position. 
Figure \ref{fig:13}C shows that the assemblies found spike at different positions along the linear track, indicating that place cells are being identified. To validate our results, we compared to the results obtained with a place cell detection technique and found that these four cell populations account for 100.0\% of the place cells (J. Taxidis, personal communication). Notably, our method only used the spike-trains to detect these cells and was able to extract also the inhibitory neurons.

\section{Discussion}
We introduce here a versatile technique to detect cell assemblies directly from spike-train data. The method uses biophysically inspired notions to create a functional connectivity matrix reflecting neuron-to-neuron relationships extracted from the observed spike dynamics. Groups of neurons are obtained from the functional connectivity matrix using a graph theoretical method for community detection, which scans partitions across all resolutions and detects relevant groupings without prescribing the level of granularity or the number and size of groups \textit{a priori}. In contrast to most other methods, our technique is able to extract hierarchical structure in recorded data; incorporates functional differences between excitatory and inhibitory neurons; and can detect the directionality emanating from feedforward connectivity.
All these are vital aspects enabling novel types of analyses in recorded datasets. 
As the method relies only on spike timings, it can be applied to both electrophysiological and optical recordings.
We tested the performance of the method on a variety of synthetic data, where we showcased its ability to extract clustered and hierarchical assemblies, in contrast to standard methodologies, such as hierarchical clustering or Modularity which have an inherently lower performance in finding such hierarchical ensembles.
We further confirmed that the directed nature of our technique allows the inference of feedforward connectivity, minimizing information loss when going from spiking data with relevant temporal ordering to a FCM.
This may open the possibility to the inference of underlying anatomical connectivities, as well as gaining insight about feedforward connections from recorded neuronal network datasets. 

We applied the framework to the analysis of spike-train simulations from several LIF network topologies (E-E clustered, E-E hierarchical and E-I clustered organizations), which result in temporally-structured network activity.
Our technique was able to identify the hierarchical structure in such simulated data without the need to rerun the analysis adjusting the parameter settings of the spike-train similarity measure.
This capability could be of interest to detect hierarchical neuronal connectivity in real systems~\citep{McGinley2013,Savic2000,Ambrosingerson1990}, as our integrated approach can identify structure at different scales without imposing strong assumptions \textit{a priori}.
In addition, the method was able to detect clusters that included both excitatory and inhibitory neurons in LIF networks.
Taking into account the functional differences of these neuron types is a distinctive feature of our methodology.
If all neurons are treated equally, irrespective of their cell type, as is commonly done, a simple split between excitatory and inhibitory neurons is often observed.
This effect is essentially due to the strong difference in firing statistics between excitatory and inhibitory neurons, although inhibitory neurons may also exhibit a range of firing-frequencies \citep{Isaacson2011,Markram2004}.
Being able to extract such structural information, and to distinguish between excitatory and inhibitory interactions where applicable, is of interest for the understanding of the functional role of cell assemblies, as these groups are likely to include both excitatory and inhibitory neurons.

We additionally showed how our framework is able to recover biological information in three sets of experimental data from mouse RGCs, salamander RGCs, and rat hippocampal data, highlighting the versatility of the method.
In the mouse RGC data, the method can assign repeated trials of the same stimulus recorded from three different neurons to the three cells without \textit{a priori} information. Salamander RGCs recorded simultaneously in response to a stimulus of randomly flickering bars were clustered into two groups displaying distinct temporal characteristics in their responses, \textit{i.e.}, transient and sustained RGCs could be distinguished. In the hippocampal recordings of a rat translocating along a linear track, the method was able to identify all the inhibitory neurons as well as four assemblies containing all the place cells associated with specific spatial information.
An advantage of our technique in this context is that it only requires spike-timing data and can thus provide complementary information and cross-validation for other techniques currently used for place cell detection.
Furthermore, the algorithm could be optimized for place cell detection to carry out additional analyses, \textit{e.g.},  examining the spiking data of rats asleep after translocation sessions; studying the effects of varying the track length; the conjunct analysis of group firings with the recorded local field potentials; or the analysis of spatially-induced firing patterns of inhibitory neurons, among others.

While previous clustering methodologies have shown good results for particular applications, we aimed here for a method with the versatility to account for the wide differences in neuronal data while simultaneously providing a simple, interpretable approach to detect cell assemblies from spiking data.
Hence our main focus was on the conceptual and generic aspects of our dynamics-based framework, rather than on fine-tuning the technical details towards more specialized applications.
Further refinements of the method are possible (or indeed desirable for specific datasets) and could lead to improved performance.
For instance, future work could be aimed at a more explicit use of neuron firing statistics, including varying firing frequencies such as during burst periods, or at the construction of functional coupling measures with more specialized biophysical couplings, including different neuron types.
Other possible extensions of the FCM similarity measure include enforcing sparsity constraints on the couplings, possibly coupled with more refined statistical assessments of the importance of individual couplings. A particularly interesting question for future work will be to consider how cell assemblies (and underlying neuronal networks) can change their group structure over time via different mechanisms such as synaptic plasticity, and how this relates to learning.  In addition, it will be interesting to study the relationship between the neuronal time scales and the Markov times that appear as optimal in our community detection.

In passing we note that in order to distinguish neuron types, all recorded neurons should ideally receive the same inputs. For example, in visual experiments a full-field stimulus should be strongly preferred over a noisy stimulus, as otherwise neurons of the same type may have uncorrelated firing due to uncorrelated inputs.
This point is applicable to all clustering algorithms and not just the one presented here.

As modern neuronal recording techniques enable simultaneous recordings of ever increasing numbers of neurons, approaching nearly entire brains \citep{Ahrens2013}, techniques for detecting spike-train communities, such as the one proposed here, will become a vital tool to provide insight into such complex data.
Collecting meaningful data from systems neuroscience experiments is the key requirement, yet being able to provide concise, intelligible representations of these recordings is just as critical in order to identify and comprehend the spatio-temporal information encoded in the data.

\subsection*{Acknowledgements}
YNB acknowledges support from the William H. Pickering Fellowship.
MTS acknowledges support from the Studienstiftung des deutschen Volkes and a Santander Mobility Award.
MTS and MB acknowledge support through a grant to MB from the Engineering and Physical Sciences Research Council (EPSRC) of the UK under the Mathematics
underpinning the Digital Economy program.
YNB, CAA, and CK thank the G. Harold \& Leila Y. Mathers Foundation.
CAA acknowledges support from the Swiss National Science Foundation (SNSF), the Human Frontier Sciences Program (HFSP).
CAA and YNB thank the National Institute of Neurological Disorders and Stroke (NINDS).
CAA and CK thank the Allen Institute for Brain Science.
We thank M. Meister and H. Asari for the retinal ganglion cell data, K. Diba for the hippocampal data, and A. Shai, J. Taxidis, E. Schomburg, and S. Mihalas for discussions.

\footnotesize
\bibliographystyle{elsarticle-harv}
\bibliography{References}

\end{document}